\documentclass[aps,rmp,twocolumn,groupedaddress]{revtex4}

\usepackage{graphicx}

\def\ensuretext{\textrm}
\newcommand{\fig}[4][ht]{\begin{figure}[#1]\begin{center}\includegraphics[scale=#2]{#3.pdf}\vspace{-0.25 cm}\caption{#4}\label{fig:#3}\end{center}\end{figure}}
\newcommand{\widefig}[4][ht]{\begin{figure*}[#1]\begin{center}\includegraphics[scale=#2]{#3.pdf}\vspace{-0.25 cm}\caption{#4}\label{fig:#3}\end{center}\end{figure*}}

\newcommand{\ket}[1]{\ensuremath{\left|{#1}\right\rangle}}
\newcommand{\isotope}[2]{\ensuremath{^{#2}}\ensuretext{#1}}
\newcommand{\Rb}[1]{\ensuremath{^{#1}}\ensuretext{Rb}}
\newcommand{\hf}[2]{\ket{#1, \; #2}}
\newcommand{\units}[1]{\ensuretext{\thinspace #1}}

\newcommand{\etal}[0]{\emph{et al.}}

\begin{document}

\title{\Rb{85} tunable-interaction Bose-Einstein condensate machine}

\author{P. A. Altin, N. P. Robins, D. D\"oring, J. E. Debs, R. Poldy, C. Figl and J. D. Close}
\affiliation{Australian Centre for Quantum Atom Optics, Australian National University, ACT 0200, Australia}
\email{paul.altin@anu.edu.au}
\homepage{http://atomlaser.anu.edu.au/}

\date{\today}

\begin{abstract}

We describe our experimental setup for creating stable Bose-Einstein condensates of \Rb{85} with tunable interparticle interactions. We use sympathetic cooling with \Rb{87} in two stages, initially in a tight Ioffe-Pritchard magnetic trap and subsequently in a weak, large-volume crossed optical dipole trap, using the $155\units{G}$ Feshbach resonance to manipulate the elastic and inelastic scattering properties of the \Rb{85} atoms. Typical \Rb{85} condensates contain $4\times10^4$ atoms with a scattering length of $a=+200\units{$a_0$}$. Our minimalist apparatus is well-suited to experiments on dual-species and spinor Rb condensates, and has several simplifications over the \Rb{85} BEC machine at JILA \cite{papp06,papp07}, which we discuss at the end of this article.

\end{abstract}

\maketitle

\section{Introduction}

In recent years there have been a number of important discoveries in condensed matter physics permitted by the ability to change the interactions between ultracold bosons. In the weakly-interacting regime, near-ideal Bose gases have recently been used to observe Anderson localisation \cite{billy08,roati08}, and promise to improve coherence times in atom interferometers \cite{gustavsson08,fattori08}. The strongly-interacting regime, on the other hand, has been used to study beyond mean-field effects, including the collapse and explosion of attracting Bose-Einstein condensates known as the `bosenova' \cite{cornish00}.

The majority of these experiments have used a Feshbach resonance, a magnetically-tunable molecular bound state, to vary the elastic scattering properties of the Bose gas. Tuning a bound state around the atomic scattering threshold using the Zeeman effect results in the divergence of the $s$-wave scattering length $a$ according to
\begin{equation}
a = a_{\text{bg}} \left( 1 - \frac{\Delta}{B-B_0} \right) \quad,
\end{equation}
where $a_{\text{bg}}$ is the scattering length far from $B_0$, the magnetic field at the resonance, and the width $\Delta$ is the difference between $B_0$ and the magnetic field at which $a=0$. Such resonances have been used to modify the $s$-wave scattering length in Bose-Einstein condensates of \isotope{Na}{23} \cite{inouye98}, \Rb{85} \cite{cornish00}, \isotope{Cs}{133} \cite{weber02}, \isotope{K}{39} \cite{roati07}, \isotope{Cr}{52} \cite{werner05,lahaye07} and \isotope{Li}{7} \cite{pollack09}. Feshbach resonance physics also led to the successful creation of a molecular BEC from a Fermi gas \cite{regal04}, and has allowed significant advances in the field of quantum chemistry \cite{kohler06}.

The first Bose-Einstein condensate (BEC) with tunable interactions was of \Rb{85}, created in the Wieman group at JILA in 2000 \cite{cornish00}. This species has a broad, low-field Feshbach resonance that allows the $s$-wave scattering length to be tuned over several orders of magnitude, including both positive and negative values (Figure \ref{fig:feshbach}). However, poor elastic and inelastic scattering cross-sections make condensation of \Rb{85} notoriously difficult to achieve. Indeed, until now only one group has successfully realized this goal.

\fig{0.45}{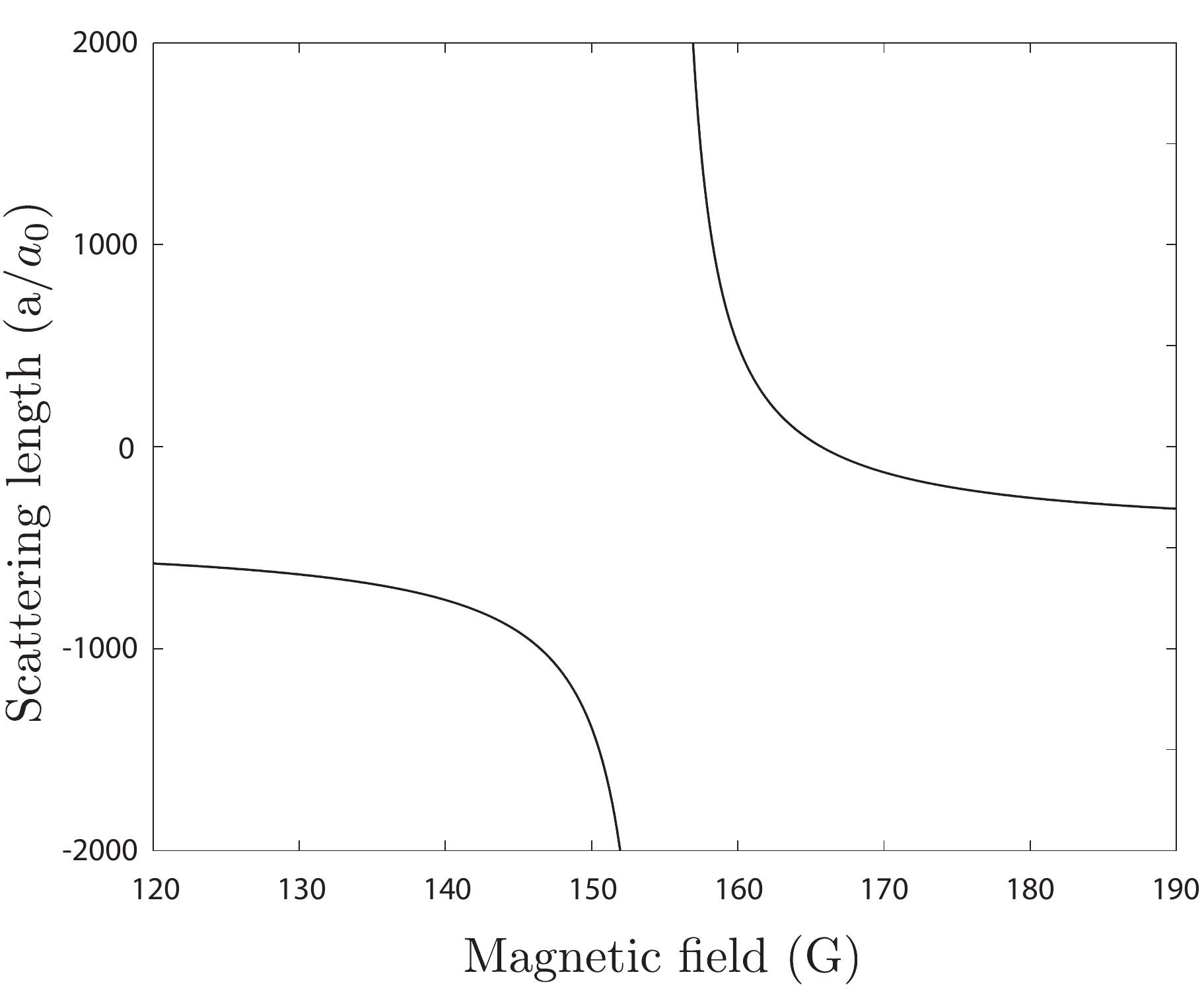}{Variation of $s$-wave scattering length with magnetic field near the \Rb{85} Feshbach resonance. The parameters of the resonance are $a_{\text{bg}}=-443\units{$a_0$}$, $B_0 = 155.0\units{G}$ and $\Delta = 10.7\units{G}$ \cite{claussen03}.}

\widefig{0.45}{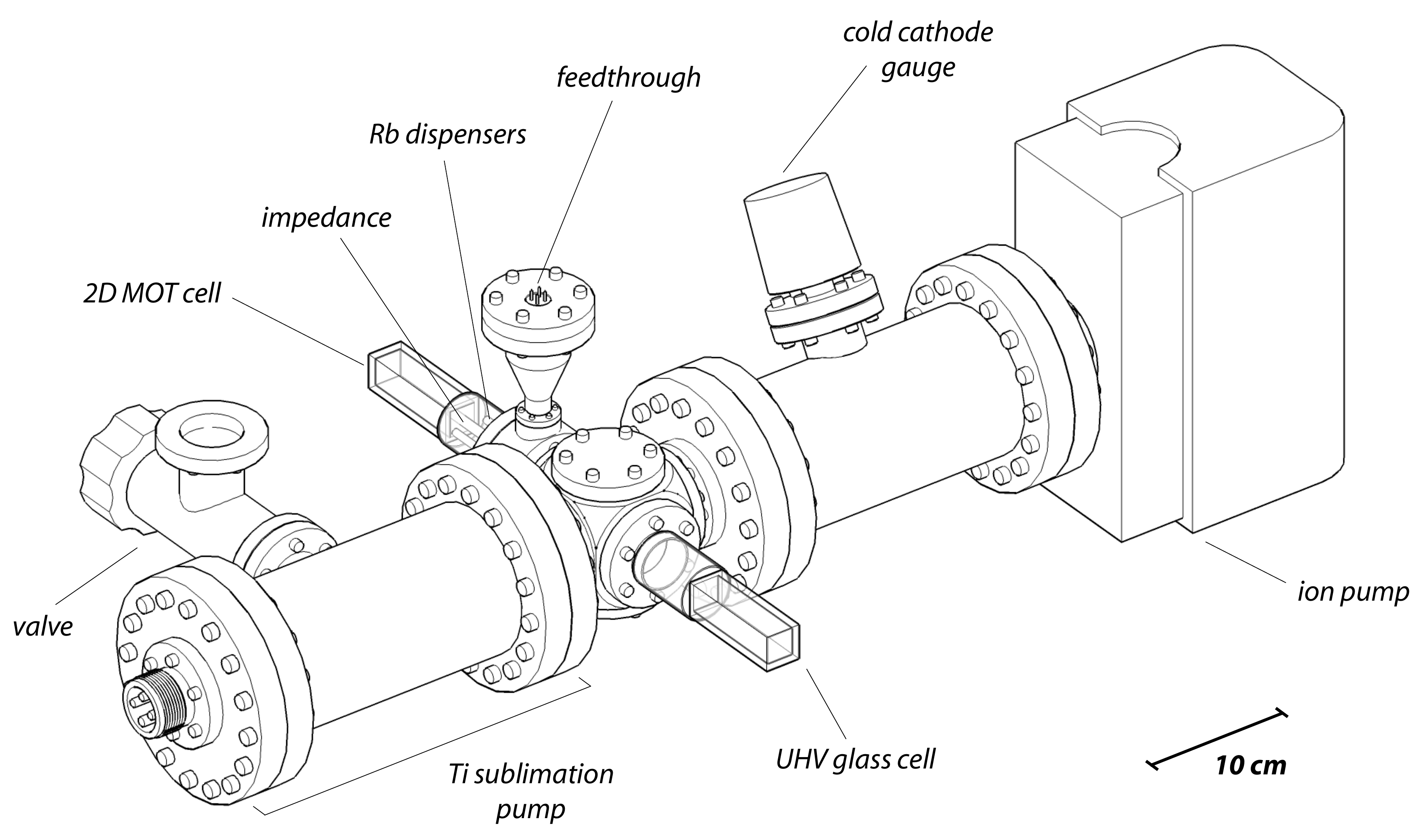}{Vacuum system schematic. The 2D and 3D MOTs are collected in coaxial quartz glass cells, connected by a small  impedance which maintains a pressure differential between the two sections.}

In this article we describe our setup for achieving Bose-Einstein condensation of \Rb{85}, based on the recent work of Papp \etal\ \cite{papp06,papp07}. We exploit the large interspecies collision cross-section between the two isotopes \cite{burke98} to sympathetically cool \Rb{85} using \Rb{87} as a refrigerant, first in a magnetic trap and subsequently in a weak, large-volume optical dipole trap. Our machine creates tunable-interaction \Rb{85} BECs containing $4\times10^4$ atoms. We have refined the approach of Papp \etal\ \cite{papp07}, resulting in a substantially simpler setup and a shorter duty cycle. A detailed comparison of the two machines is presented at the end of this paper.

\section{Overview}

In our system, a cold beam containing both \Rb{85} and \Rb{87} atoms is produced in a two-dimensional magneto-optical trap (MOT) and is directed through the vacuum system to a retroreflected, three-dimensional MOT in the main chamber, which is evacuated to a pressure of $\sim10^{-11}\units{Torr}$. The atoms are pumped into their lower ground states before being loaded into a quadrupole Ioffe-Pritchard (QUIC) magnetic trap. Here, the \Rb{87} is selectively evaporated by a radiofrequency sweep which drives transitions to untrapped states, sympathetically cooling the \Rb{85} through thermal contact. Once the temperature of the combined sample has fallen to $20\units{$\mu$K}$, the atoms are transferred to a crossed optical dipole trap and a large magnetic bias field is applied to suppress inelastic collisions in \Rb{85}. Finally, the depth of the dipole trap is reduced, resulting in further evaporation of both species. With the appropriate magnetic field strength, the \Rb{85} scattering length can be made positive and a stable condensate of $4\times10^4$ atoms is created.

The use of an optical dipole trap to confine the sample at low temperatures facilitates control of the scattering properties of \Rb{85} using the magnetic Feshbach resonance at $155\units{G}$. However, direct loading of the optical trap from the MOT is not feasible due to large density-dependent inelastic losses in \Rb{85}. Thus we require pre-cooling of the sample in a magnetic potential before transfer to the dipole trap for final evaporation. In our system the QUIC trap, the dipole trap and the Feshbach field coils are all co-axial, which facilitates transfer between the traps and control of the bias field. We image the atoms through absorption of resonant light.

\section{Vacuum system}

A schematic of our vacuum system is shown in Figure \ref{fig:vacuumschematic}. The setup is simple and compact compared with many BEC machines. Two $25\units{mm} \times 25\units{mm} \times 80\units{mm}$ quartz glass cells for the two- and three-dimensional MOTs are flanked by a $75\units{l/s}$ ion pump and a titanium sublimation pump housed in a $6$-inch tube. An electrical feedthrough provides current to two $50\units{mg}$ rubidium dispensers (Alvatec AS-Rb-50-F) mounted in the 2D MOT cell. Pressure readings are taken from the ion pump and from a cold cathode gauge (Pfeiffer IKR 270) mounted on one of the large tubes, and agree to within a factor of 2 in their specified operating ranges. After construction, the entire system was baked at $300^\circ\text{C}$ for six days.

To create a BEC, an ultra-high vacuum of $\lesssim 10^{-10}\units{Torr}$ is required to reduce atom loss by collisions with background vapor particles, in order to enable efficient evaporative cooling. However, to collect a large MOT a relatively high pressure is optimal \textendash\ up to Rb vapor pressure $\sim10^{-7}\units{Torr}$. To fulfill both these requirements, our vacuum system is partitioned into two sections by a small impedance \textendash\ a stainless steel tube protruding into the 2D MOT glass cell with a $0.8\units{mm}$ diameter hole. The low conductance $C=7\times10^{-3}\units{l/s}$ of this tube allows the pressure in the two sections to differ by several orders of magnitude. A two-dimensional MOT in the high vacuum (HV) section produces a beam of cold atoms which are pushed through the impedance to feed the three-dimensional MOT in the ultra-high vacuum (UHV) cell. The pressure in the UHV cell is maintained at $3\times10^{-11}\units{Torr}$.

\section{Laser system}

The laser light used for the \Rb{85} and \Rb{87} magneto-optical traps and for absorption imaging of the atom clouds is derived from six homebuilt external cavity diode lasers (ECDLs) in Littrow configuration. These use feedback from a low-efficiency holographic diffraction grating to narrow the linewidth of a $780\units{nm}$ laser diode (Roithner ADL-78901TX), giving between $50\units{mW}$ and $80\units{mW}$ of output at the frequency of the rubidium $\text{D}_2$ line ($5^2\text{S}_{1/2} \rightarrow 5^2\text{P}_{3/2}$). The lasers are locked using saturated absorption spectroscopy and either current or Zeeman modulation,\footnote{We have found that producing error signals by current modulation can dramatically broaden the linewidth of our lasers to more than the natural linewidth of the atomic transition ($6\units{MHz}$). For this reason, we use Zeeman modulation for the lasers from which we derive imaging light, to enable maximum absorption when imaging dilute clouds on resonance.} and double-passed through acousto-optic modulators which allow the light to be shifted to the desired frequency and swiftly shuttered. The light from two of these lasers (those driving the \Rb{87} MOTs) is amplified by $1.5\units{W}$ tapered amplifier chips (m2k laser TA-780-1500). All the light is guided to the experiment in single-mode polarization-maintaining optical fibres, which serve to isolate the experiment from the laser table and to mode-clean the light. Each fibre carries two frequencies \textendash\ one for each species \textendash\ with orthogonal polarization.

\begin{table}[b] \begin{center} \resizebox{8.5cm}{!} {\begin{tabular}{lccc}
\hline\hline
\qquad Purpose & \hspace{1cm}Transition\hspace{1cm} & \hspace{0.3cm}Detuning\hspace{0.3cm} & \hspace{0.3cm}Power\hspace{0.3cm} \\
 & & (MHz) & (mW) \\
\hline
\multicolumn{4}{l}{\Rb{87}} \\
\qquad 3D MOT trapping & $F=2\rightarrow F^\prime=3$ & $-24$ & $60$ \\
\qquad 3D MOT repumping & $F=1\rightarrow F^\prime=2$ & $0$ & $20$ \\
\qquad 2D MOT trapping & $F=2\rightarrow F^\prime=3$ & $-12$ & $150$ \\
\qquad 2D MOT repumping & $F=1\rightarrow F^\prime=2$ & $0$ & $20$ \\
\qquad Push beam & $F=2\rightarrow F^\prime=3$ & $+5$ & $0.15$ \\
\qquad Imaging & $F=2\rightarrow F^\prime=3$ & $-6$ & $0.4$ \\
\\
\multicolumn{4}{l}{\Rb{85}} \\
\qquad 3D MOT trapping & $F=3\rightarrow F^\prime=4$ & $-20$ & $15$ \\
\qquad 3D MOT repumping & $F=2\rightarrow F^\prime=3$ & $0$ & $2.5$ \\
\qquad 2D MOT trapping & $F=3\rightarrow F^\prime=4$ & $-12$ & $35$ \\
\qquad 2D MOT repumping & $F=2\rightarrow F^\prime=3$ & $0$ & $3.0$ \\
\qquad Push beam & $F=2\rightarrow F^\prime=3$ & $+29$ & $0.03$ \\
\qquad Imaging & $F=2\rightarrow F^\prime=3$ & $0$ & $0.4$ \\
\hline\hline
\end{tabular}} \end{center} \vspace{-10pt}
\caption{The laser frequencies used in this experiment. Typical values for the power available after fibre coupling and the detuning relative to the specified transition are given in the columns on the right.} \label{tab:lasers}
\vspace{10pt} \end{table}

\section{Magneto-optical traps}

In a typical experimental run, $10^{10}$ \Rb{87} atoms and $10^8$ \Rb{85} atoms are collected in the 3D MOT. A low power push beam slightly detuned from the cooling transition drives atoms from the 2D MOT through the impedance into the UHV section of the vacuum system, where they are captured in the 3D MOT. Each MOT uses retroreflected beams apertured to $20\units{mm}$ in diameter. The trapping lasers for the 3D MOT are detuned $24\units{MHz}$ and $20\units{MHz}$ to the red of the $\Rb{87}$ $\ket{F=2}\rightarrow\ket{F^\prime=3}$ transition and the $\Rb{85} \ket{F=3}\rightarrow\ket{F^\prime=4}$ transition respectively. The 2D MOT trapping lasers are each tuned 12\units{MHz} to the red of these transitions. To prevent loss of atoms through off-resonant transitions to the lower ground states, which occur roughly once in $10^4$ absorption events, repumping light resonant with the $\Rb{87}$ $\ket{F=1}\rightarrow\ket{F^\prime=2}$ transition and the $\Rb{85} \ket{F=2}\rightarrow\ket{F^\prime=3}$ transition is mixed in with the trapping light in both MOTs.

We have also experimented with using a single ECDL locked to the \Rb{87} cooling transition to drive both 2D MOTs by passing the light through an electro-optic modulator (EOM) in a microwave cavity driven at $1.1265\units{GHz}$ \textendash\ corresponding to the frequency difference between the cooling transitions for each isotope. Up to $20\%$ of the total power could thus be transferred into a sideband resonant with the \Rb{85} cooling transition (a further $20\%$ in the upper sideband was not resonant with either species). This light was then amplified by a $1.5\units{W}$ tapered amplifier chip; $250\units{mW}$ was available at the experiment after fibre-coupling. Although this setup was easier to align (since the 2D MOTs were automatically coincident) and allowed us to produce larger \Rb{85} MOTs, we found that thermal and mechanical instability in the EOM caused large run-to-run atom number fluctuations.

Using either approach, the flux from the 2D MOT is sufficient to fill the 3D MOT within $5\units{s}$. After a $20\units{ms}$ polarization-gradient cooling stage, the MOT repumping lasers are switched off for $1\units{ms}$ to allow both species to be optically pumped into their lower ground states (\Rb{87} \ket{F=1} and \Rb{85} \ket{F=2}) in preparation for sympathetic cooling.

\section{Magnetic trap}

Once the MOT light has been switched off, the low field seeking states \Rb{85} \hf{F=2}{m_F=-2} and \Rb{87} \hf{F=1}{m_F=-1}  are captured in the quadrupole magnetic field produced by the MOT coils and then magnetically transported over a distance of $40\units{mm}$ to a second quadrupole trap with the aid of a rectangular transfer coil (see Figure \ref{fig:trapcoils}). In this process, the current in the transfer coil is ramped up to push the cloud towards the Ioffe-Pritchard trap coils, then the MOT and transfer coils are turned off as the current in the quadrupole coils ramps up. This quadrupole trap is then converted into a harmonic Ioffe-Pritchard potential using the quadrupole-Ioffe coil configuration (QUIC) \cite{esslinger98}.

\fig{0.32}{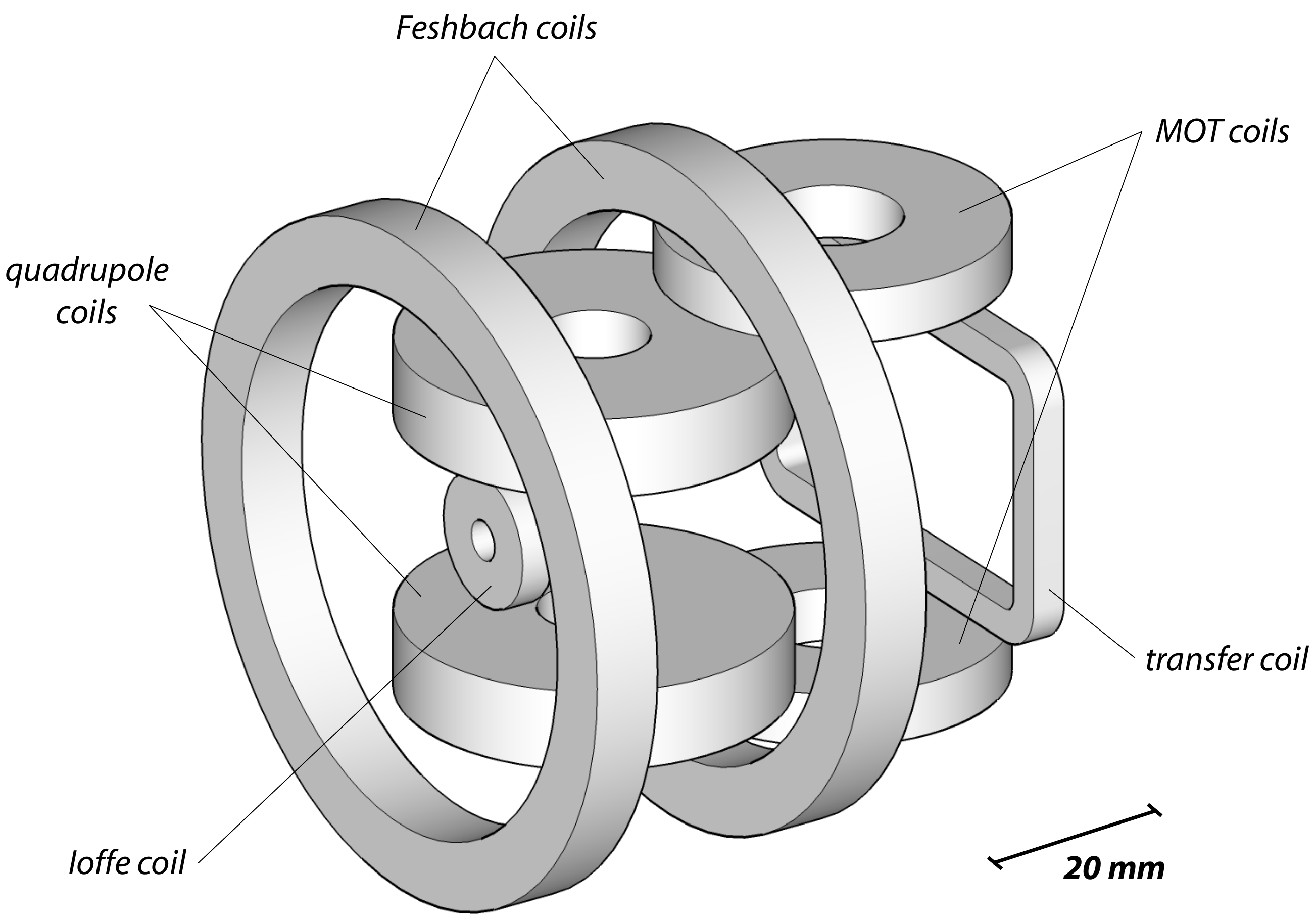}{Magnetic trap coils mounted around the UHV cell. Not shown are the rf coil, which produces the oscillating magnetic field used for evaporating atoms from the magnetic trap; and the imaging coil, which defines a quantisation axis to ensure maximum absorption of the probe light during imaging.}

The QUIC trap coils were designed using a full numerical optimisation of the trapping frequencies at a given power. The calculation was performed taking into account the various geometric constraints of the experiment and provided the optimal size and position of the coils, the diameter of the wire to be used for the windings, and the current to flow in each coil. The quadrupole coils each consist of $137$ turns of enamelled copper wire with diameter $1.35\units{mm}$ and run at $25\units{A}$, dissipating a total power of $270\units{W}$. The Ioffe coil has 33 turns of $0.9\units{mm}$ wire and also runs at $25\units{A}$, dissipating $40\units{W}$.

The trap coils are cooled by a recirculating solid state chiller, which pumps water at $13^\circ\text{C}$ through holes in the aluminium trap mount. To maximize the cooling efficiency, the coils were wound directly onto the mount using a high thermal conductivity epoxy (Cotronics Duralco 132). Before winding the Ioffe coil, we mixed fine (filed) copper dust with the epoxy in roughly a $1:2$ ratio to improve its thermal conductivity ($k_{\text{Cu}} = 380\units{W/m$\cdot$K}$ compared to $k_{\text{epoxy}} = 6\units{W/m$\cdot$K}$). This had no detrimental effect on the properties of the epoxy.

Since we do not use the magnetic trap to study condensates, there is no stringent requirement on the stability of the bias field, thus we can use separate power supplies to drive the quadrupole and Ioffe coils. This facilitates small adjustments of the bias field. Fast switch-off of the trap coils is essential for absorption imaging of the atoms in the magnetic trap and when transferring samples to the dipole trap. We are able to achieve a switching time of $200\units{$\mu$s}$ using a solid-state relay with a diode/resistor shunt around the coils. Radial slots cut into the aluminium coil mount also help by minimizing eddy currents during switch-off.

The final magnetic potential has trapping frequencies of $\omega_z = 2\pi\times16\units{Hz}$ axially and $\omega_\rho = 2\pi\times156\units{Hz}$ radially for \Rb{87} atoms in the \hf{F=1}{m_F=-1} hyperfine state, and a bias field of $3.7\units{G}$. Typically, $3\times10^9$ \Rb{87} atoms and $1\times10^7$ \Rb{85} atoms are present at $200\units{$\mu$K}$ in the QUIC trap before evaporation. The lifetime of the atoms in the magnetic trap is measured to be $20\units{s}$, resulting in a background loss of roughly $50\%$ during the $15\units{s}$ of rf evaporation. This loss is due to collisions with background gas particles, predominantly rubidium vapor from the 2D MOT chamber. The lifetime can be increased by lowering the pressure in the 2D MOT cell, however this reduces both the rate at which the 3D MOT fills and the final number of atoms collected. The pressure is controlled by the current flowing through the dispensers, and is chosen to maximize the number of atoms present at the end of the evaporation.

Radiofrequency forced evaporation is used to cool the samples in the magnetic trap to $\mu$K temperatures. An oscillating rf field induces magnetic dipole transitions between trapped and untrapped spin states to expel the most energetic atoms, and those remaining re-thermalize via elastic collisions. We use an arbitrary waveform generator (Agilent 33250A) and a $4\units{W}$ rf amplifier (Mini-Circuits TIA-1000-1R8) to drive a two-loop coil of radius $12\units{mm}$ placed against the glass cell. The function generator produces a logarithmic rf sweep from $50\units{MHz}$ to $\sim3\units{MHz}$ over $15\units{s}$, with the final frequency determining the temperature of the sample. In the absence of \Rb{85} atoms, we are able to produce pure \Rb{87} \hf{F=1}{m_F=-1} condensates containing $2\times10^6$ atoms in the QUIC trap after $15\units{s}$ of evaporation.

\section{Control and Imaging}

Our machine is controlled by a desktop computer running National Instruments LabVIEW 7. The computer is equipped with two 8-channel 16-bit analog output cards (NI PCI-6733), giving variable voltage output of $-10\units{V}$ to $+10\units{V}$, and a 32-channel digital card (NI PCI-6533) with $+5\units{V}$ TTL outputs. A computer clock signal updates the state of each channel, providing timing control accurate to $< 100\units{ns}$ with a resolution of $100\units{$\mu$s}$. During a long run, the memory required to store the state of all analog channels at each update can exceed the capacity of the card's buffer. To overcome this difficulty, two of the digital channels are reserved for triggering analog updates only when a channel value is altered. Since there are often long periods during a run when no updates are required (e.g. during MOT loading), this substantially reduces the amount of information that must be stored in the buffer.

Absorption imaging is our primary diagnostic tool for ultracold atom clouds. A $100\units{$\mu$s}$ pulse of repumping light is applied first to pump the atoms out of the lower ground state, immediately followed by a $100\units{$\mu$s}$ pulse of light on the cooling transition. The imaging light is circularly polarized and a small magnetic bias field is applied along the direction of the probe light to ensure that the imaging transition is closed. The shadow left by the atoms on the imaging laser is captured with an IEEE1394 CCD camera (Point Grey Research Dragonfly) and acquired using Coriander for Linux \cite{coriander07} on a separate computer. The images are processed using MATLAB to determine the number, spatial distribution and temperature of either species at the end of a run.

\section{Sympathetic cooling}

Sympathetic pre-cooling in the magnetic trap is critical to achieving \Rb{85} BEC in our setup. This is because achieving a sufficiently high phase space density in an optical trap loaded directly from a MOT would either involve very high atomic densities, for which density-dependent inelastic losses would prevent efficient evaporation; or would require a prohibitively large amount of laser power. Pre-cooling in a magnetic trap allows us to create cold samples with phase space densities on the order of $10^{-3}$, which can then be loaded into a shallow, large-volume optical dipole trap for final evaporation to quantum degeneracy.

With both species in the QUIC trap, rf evaporation predominantly removes \Rb{87} \hf{F=1}{m_F=-1} atoms, cooling both species as long as they remain in thermal contact. The reason for the isotope selectivity is twofold: firstly, since the Land\'e factor $g_F$ has a larger magnitude for the \Rb{87} ground state ($|g_F| = 1/2$) than for \Rb{85} ($|g_F| = 1/3$), the \Rb{87} evaporation surface at a given radiofrequency is closer to the centre of the trap than for \Rb{85}; and secondly, the \Rb{85} \hf{F=2}{m_F=-2} cloud is more tightly confined in the magnetic trap (i.e. smaller at a given temperature) due to its larger $g_F\;m_F$ factor. The interspecies \Rb{87}-\Rb{85} collision cross-section is large in the temperature range of interest (below $200\units{$\mu$K}$) \cite{burke99}, and the two clouds remain spatially overlapped until the temperature drops below $\sim1\units{$\mu$K}$, where the difference in the gravitational sag for each species becomes comparable to the size of the clouds. The present limit to this pre-cooling stage is therefore only the number of \Rb{87} atoms initially present in the magnetic trap: although we collect enough to create pure \Rb{87} condensates, the addition of \Rb{85} places a heat load on the sample which prevents us from reaching sub-$\mu$K temperatures in \Rb{85}.

Figure \ref{fig:sympathetic} shows the number of atoms as a function of temperature for each species during the magnetic trap pre-cooling. The single-species \Rb{87} evaporation is also shown. The isotope selectivity of the evaporation is clearly demonstrated by the fact that no loss of \Rb{85} is detected during cooling from $200\units{$\mu$K}$ to $20\units{$\mu$K}$, corresponding to an increase in phase space density of over three orders of magnitude. The sympathetic cooling trajectory of \Rb{85} begins to roll off at around $20\units{$\mu$K}$, as the number of each species present becomes comparable. This is to be expected, since beyond this point the density of \Rb{85} at the evaporation surface will not necessarily be lower than that of \Rb{87}, and so the rf begins to remove both species from the trap. 

\fig[t]{0.58}{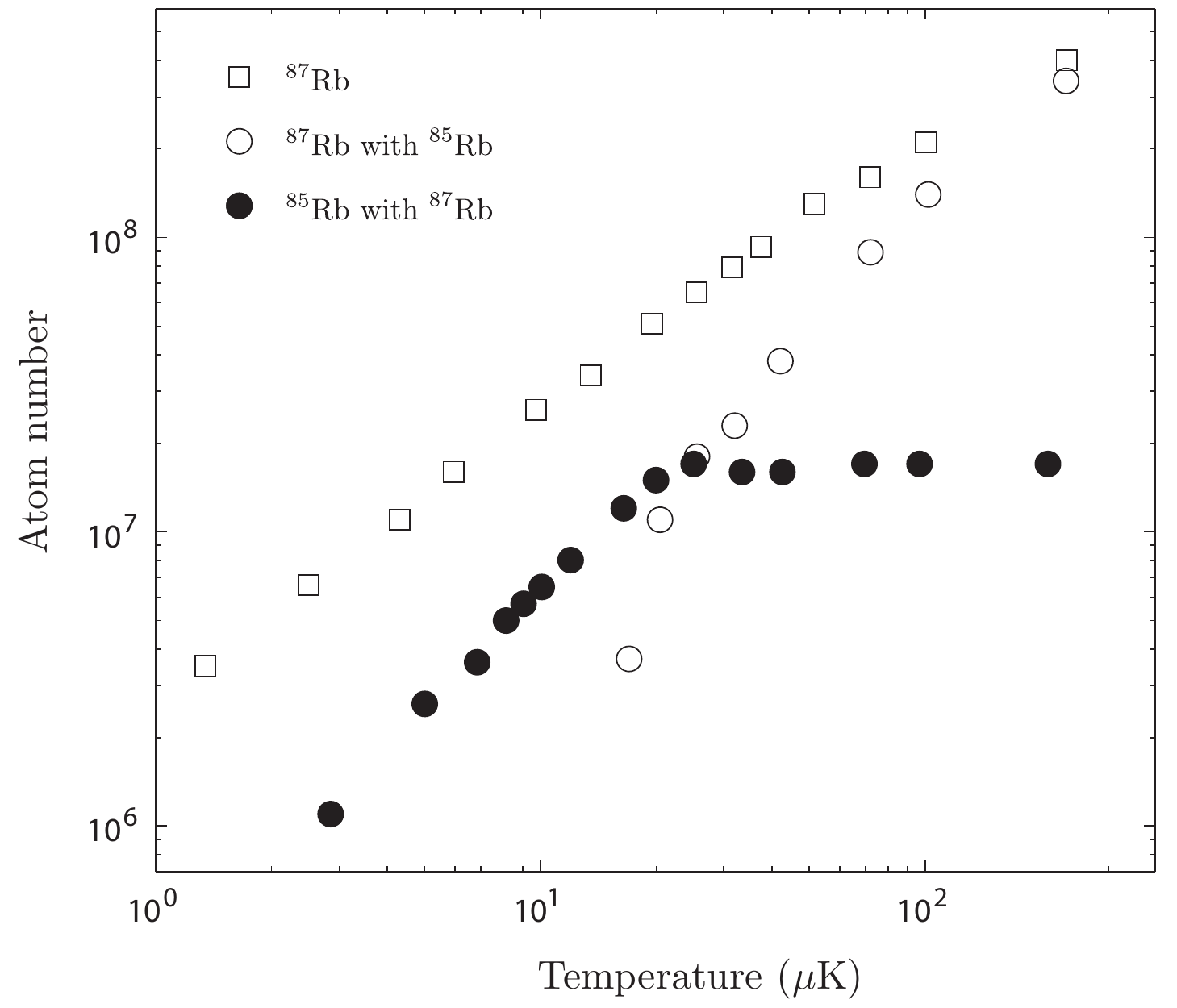}{Atom number as a function of temperature during rf-induced evaporative pre-cooling in the magnetic trap. The evaporation trajectories of \Rb{85} (filled circles) and \Rb{87} (open circles) are shown, as well as the single-species trajectory for \Rb{87} (squares).}

Under optimal experimental conditions, we can create samples of $8\times10^6$ \Rb{85} atoms at $10\units{$\mu$K}$ in the QUIC trap, with a phase space density of $6\times10^{-4}$. When making BEC, we typically begin with a smaller sample of \Rb{85} in order to have more \Rb{87} remaining in the trap, and cease rf evaporation at $20\units{$\mu$K}$ with $6\times10^6$ \Rb{85} atoms and $4\times10^7$ \Rb{87} atoms available for transfer to the optical dipole trap.

\section{Optical dipole trap}

In order to create and study a \Rb{85} Bose-Einstein condensate with more than $\sim100$ atoms, it is necessary for the atoms to be in a magnetic field at which the $s$-wave scattering length $a$ is positive. The most convenient point at which this occurs is on the high-field side of the $155\units{G}$ Feshbach resonance. However, it is difficult to make a tightly-confining magnetic trap with such a large bias field $B_0$, since for standard Ioffe-Pritchard-type traps the radial trapping frequency falls as $\omega_\rho \propto 1/\sqrt{B_0}$. Furthermore, many of the experiments one would like to perform on a \Rb{85} condensate involve changing the scattering length via the magnetic field, and the relationship between $B_0$ and $\omega_\rho$ means that bias field adjustments would couple into changes in radial confinement, which could cause undesirable excitations in the sample. For this reason, we confine the atoms in a far-detuned crossed-beam optical dipole trap for the final stage of the experiment and use a set of dedicated `Feshbach' coils to generate a homogeneous magnetic field for manipulation of the scattering length. The light for the dipole trap is sourced from a $20\units{W}$ Er-doped fibre laser (SPI Lasers redPOWER compact) operating at $1090\units{nm}$. The laser beam has a diameter of $5\units{mm}$, an M$^2$ value of $<1.1$ and an emission linewidth of $2\units{nm}$ FWHM.

The optical setup for the dipole trap (Figure \ref{fig:dipoletrap}) is remarkably simple. The randomly-polarized output from the fibre laser is split evenly on a $1$-inch polarising beamsplitter. Each beam then passes through a $2$-inch plano-convex lens with focal length $f=100\units{cm}$. The axial beam is directed into the end of the glass cell through the centre of the Ioffe coil, parallel to the long axis of the QUIC trap. The cross-beam also passes through the cell in the horizontal plane, making an angle of $75^\circ$ with the axial beam (see Figure \ref{fig:dipoletrap}). The intensity of the dipole trap is controlled by a $0-10\units{V}$ analog input on the laser controller. In contrast to most optical BEC experiments, we do not require any active stabilisation of the laser intensity. In addition, the short coherence length ($<1\units{mm}$) of the laser prevents standing wave patterns forming in the overlap region, eliminating the need for the frequency of the cross-beam to be shifted using an acousto-optic modulator, which would reduce the power available for trapping.

\fig{0.42}{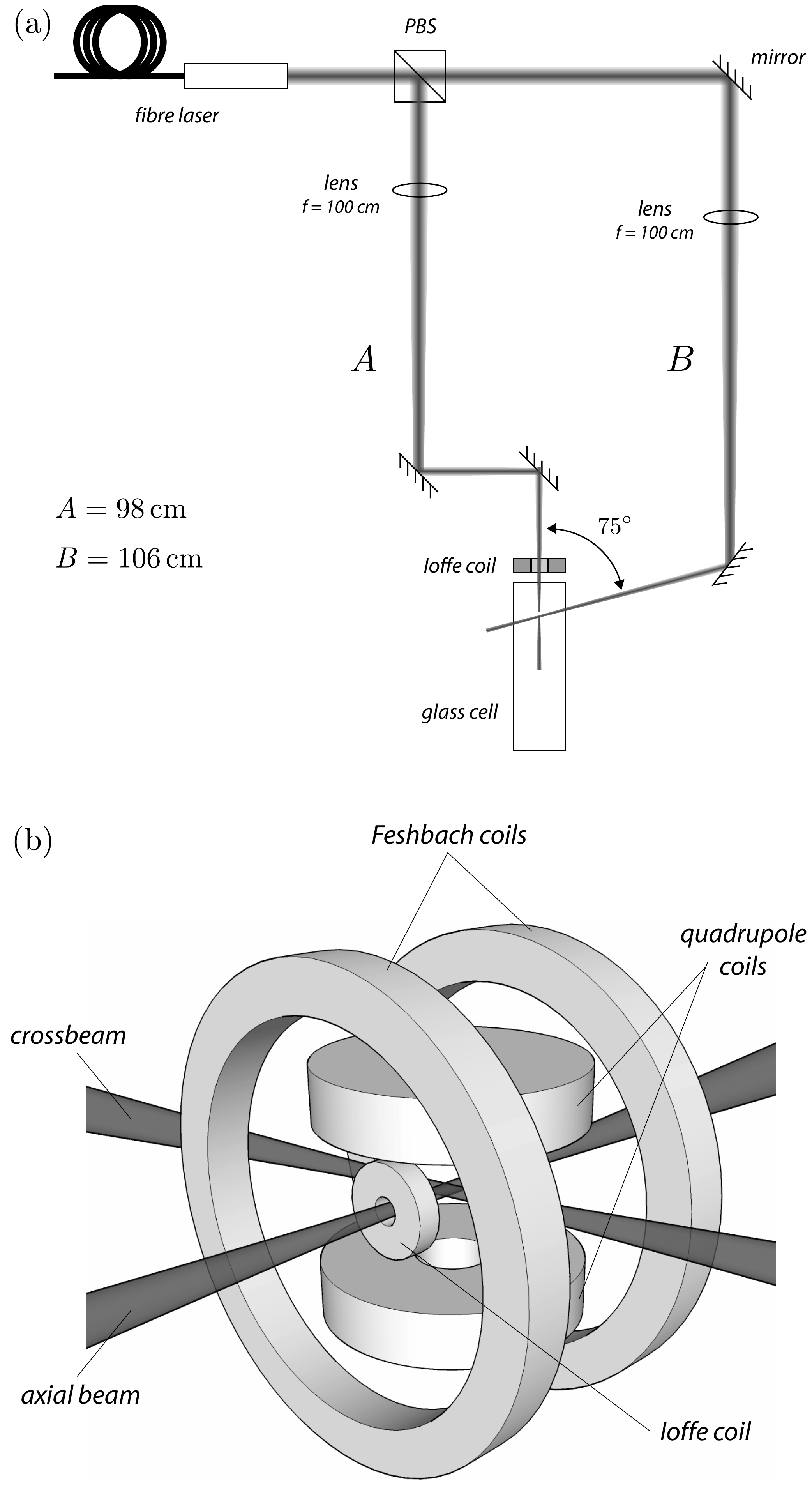}{$(a)$ Schematic diagram of the optical setup for the dipole trap. $A$ and $B$ are the paths taken by the axial and cross-beam respectively, which cross in the cell at an angle of $75^\circ$. $(b)$ The position of the dipole beams and Feshbach coils in relation to the magnetic trap.}

The lens in the axial (cross) beam is placed $98\units{cm}$ ($106\units{cm}$) from the atoms, so that the beam radius at the cloud is approximately $150\units{$\mu$m}$ ($200\units{$\mu$m}$). Although the cloud is not precisely at the waist of either beam, the long Rayleigh range $z_R>5\units{cm}$ and resulting weak trapping force along each beam (equivalent to $<1\units{Hz}$) means that the trap profile in each direction is dominated by the perpendicular beam. At full power, with around $9\units{W}$ in each beam, the trap is approximately $30\units{$\mu$K}$ deep \textendash\ corresponding to roughly half of the effective depth of the magnetic trap at the final frequency of the rf evaporation. The radial trapping frequency is $\omega_\rho=2\pi\times110\units{Hz}$.

Following the sympathetic pre-cooling stage, the dipole beams are suddenly (within $<1\units{ms}$) superimposed onto the trapped cloud. If the magnetic field is switched off, all of the atoms remain trapped radially, but spread out along the axial beam since the cross-beam is not large enough in this direction to contain the entire cloud, which is about $2\units{mm}$ long. For this reason, we leave the QUIC trap coils running at $50\%$ to create a magnetic field curvature corresponding to approximately $11\units{Hz}$ along the axis. After further evaporation, the cloud is small enough to fit into the cross-beam and this field can be switched off. The radial confinement of the magnetic trap is removed automatically once the Feshbach magnetic field is applied, as described below.

\section{Feshbach magnetic field}

\widefig{0.9}{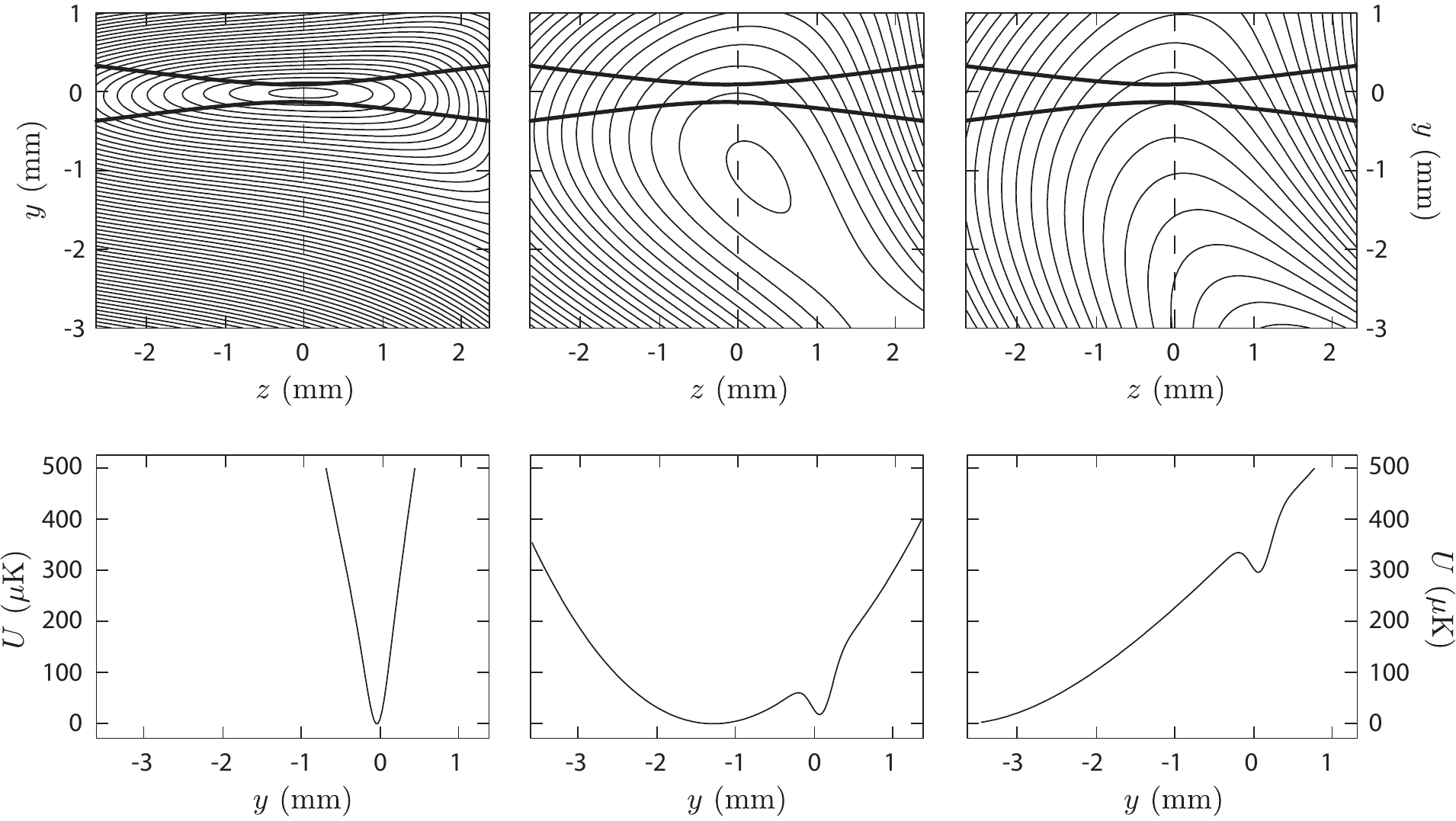}{Splitting of the cloud during the ramp up of the bias field, with the current in the Feshbach coils at $0\%$, $50\%$ and $100\%$. The upper panel shows contour lines of the magnetic plus gravitational potential, with the axial dipole trapping beam overlaid (not to scale). The lower panel shows the shape of the total potential seen by the atoms in the $y$ direction (along the dashed lines on the contour plots). The formation of a double well splits the trapped cloud in two, and those atoms not confined by the optical trap are lost as the magnetic trap sags under gravity and eventually reaches the edge of the glass cell.}

Manipulating the scattering properties of \Rb{85} using the $155\units{G}$ Feshbach resonance is crucial in the final stages of cooling toward BEC, for two reasons: firstly, a BEC of significant size will not be stable far from the Feshbach resonance due to the negative background $s$-wave scattering length $a_{\text{bg}}=-443\units{$a_0$}$; and secondly, as the cloud cools and its density increases, inelastic losses become increasingly severe. Although our pre-cooling stage is at present limited by the number of refrigerant \Rb{87} atoms in the magnetic trap, it has been shown that inelastic collisions preclude efficient cooling of \Rb{85} below about $5\units{$\mu$K}$ at low field \cite{papp07}. Both of these issues may be overcome by tuning the scattering cross-sections using the Feshbach resonance.

The coils we use to apply the Feshbach magnetic field are wound from the same wire as the quadrupole coils, and are mounted in the same water-cooled aluminium block for mechanical and thermal stability. Each coil consists of $55$ turns and runs at $13-15\units{A}$ to produce a field in the range $150-170\units{G}$. The coils are in Helmholtz configuration, producing a field parallel to that of the QUIC trap, and are centred on the QUIC trap minimum, which is $5\units{mm}$ from the centre of the quadrupole coils. This ensures optimal field uniformity over the trapped cloud: the field gradient at the atoms is well below $1\units{mG/cm}$, and the cloud at $20\units{$\mu$K}$ is only $2\units{mm}$ in length.

Immediately after the optical dipole trap is switched on, the current in the Feshbach coils is ramped up over $500\units{ms}$ to increase the magnetic bias field to $168\units{G}$. Since the radial curvature of the QUIC trap (which is now running at $50\%$) is inversely related to the bias field, this relaxes the radial confinement of the magnetic trap to the point where it is insufficient to support the atom cloud against gravity. Once at the full bias field, therefore, the atoms are held in a kind of hybrid trap, with the radial confinement provided purely by the axial dipole trapping beam and the axial confinement provided by the magnetic field curvature of the weakened QUIC trap.

An unfortunate side effect of this process is the splitting of the cloud by the resulting double-well potential as the bias field is ramped up. This effect is illustrated in Figure \ref{fig:splitting}. Initially, the atoms sit at the minimum of the potential created by the magnetic field and gravity (contour plot), and it is to this position that the optical dipole trap is aligned. However, as the bias field increases, the radial magnetic curvature decreases as described above, and so the minimum of the magnetic plus gravitational potential moves downward (the gravitational sag is proportional to $g/\omega_\rho^2$). This creates two distinct minima in the total potential seen by the atoms, and as they separate, the cloud is split.

In addition to those lost by the splitting mechanism described above, some atoms leave as the bias field is ramped up because the dipole trap is shallower than the magnetic trap at the end of the rf evaporation. This can itself be considered evaporation, since it is only the most energetic atoms that will spill over the edge of the trap. The combination of these two effects causes a loss of about $50\%$ in number for each species, but keeps the phase space density of the sample constant during the transfer from QUIC trap to optical trap. We begin the final evaporation with $3\times10^6$ \Rb{85} atoms and $2\times10^7$ \Rb{87} atoms in the dipole trap at a temperature of $5\units{$\mu$K}$ and a phase space density of $10^{-3}$.

Before attempting evaporation to BEC in the optical trap, it is important to know the magnetic field produced by the Feshbach coils precisely. We calibrate the field by addressing the atoms with radiofrequency to drive transitions between the $m_F$ sublevels. At the fields of interest ($>100\units{G}$), the nonlinear contribution to the Zeeman splitting is large compared to the frequency width of the cloud, so that the transitions are well separated from each other and only two states are coupled by the rf.

There are a number of ways in which transitions between $m_F$ states can be observed. With either species, the substates can be spatially separated by applying a magnetic field gradient during the expansion, and then imaged separately. Alternatively, using \Rb{87} the trapped \hf{F=1}{m_F=-1} state can be coupled to the untrapped \hf{F=1}{m_F=0} state. These atoms no longer feel the axial magnetic confinement and leave the trap along the axial beam, resulting in observable loss from the cloud. Interestingly, we also observe loss when coupling the \hf{F=2}{m_F=-2} and \hf{F=2}{m_F=-1} states in \Rb{85}, despite the fact that both are low-field seeking and should remain in the trapping region. We attribute this to inelastic collisions in the \ket{m_F=-1} state, which occur at a rate almost 6 times higher than in the \ket{m_F=-2} state \cite{altin09}.

Once the frequency which drives transitions between neighboring Zeeman states is known, the Breit-Rabi equation \cite{breit31} can be used to calculate the field at the centre of the cloud. Using this technique we are able to calibrate the Feshbach bias field to within $50\units{mG}$.

\section{Bose-Einstein condensation}

\fig[t!]{0.35}{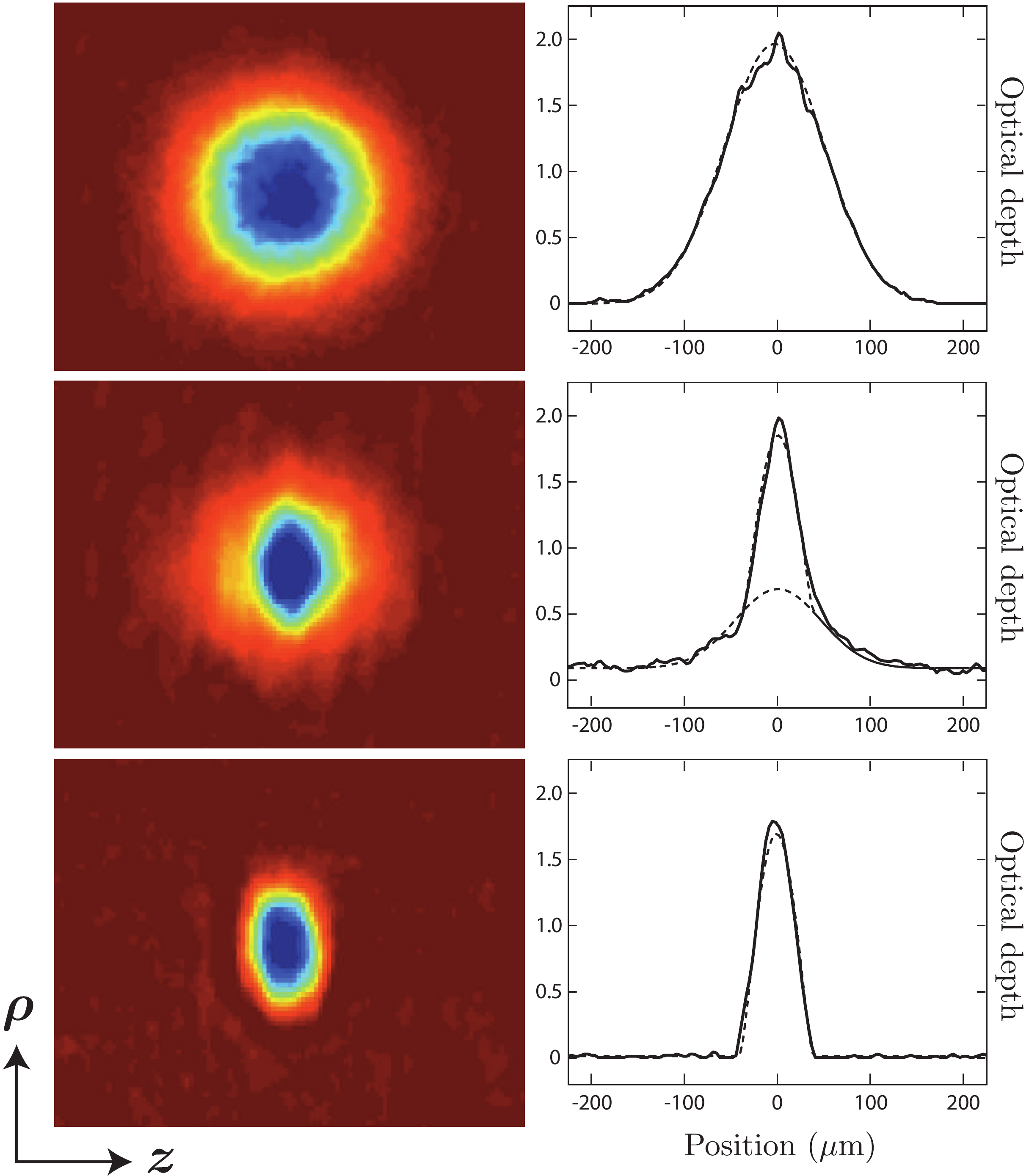}{$(a)$ Absorption images taken after $20\units{ms}$ of ballistic expansion showing the formation of a \Rb{85} BEC as the depth of the crossed dipole trap is reduced. The elongation of the cloud along the tight trapping direction demonstrates the reversal of aspect ratio characteristic of BEC. $(b)$ Optical depth profiles along the $z$ direction, with fits showing the transition from a thermal Gaussian distribution, to a bimodal distribution, to a pure condensate.}

After single-species pre-cooling in the QUIC trap, we are able to produce pure \Rb{87} \hf{F=1}{m_F=-1} condensates containing $2\times10^6$ atoms in the QUIC trap after $7\units{s}$ of evaporation in the optical dipole trap. The strength of the Feshbach bias field is not important in this case, and using Landau-Zener sweeps to effect rapid adiabatic passage between the Zeeman states we can create and study the behavior of spinor \Rb{87} \ket{F=1} condensates as a function of magnetic field.

Inelastic losses in \Rb{85} \hf{F=2}{m_F=-2} are minimized at a magnetic field of $168\units{G}$ \cite{roberts00,altin09}. With the Feshbach coils set to this field, we proceed to evaporate the sample further by reducing the power in the dipole trap. This lowers the depth of the trap and allows the most energetic atoms to spill out. With the axial magnetic field curvature still present, the evaporation is only two-dimensional (atoms leave the trap only radially). At $168\units{G}$ the $s$-wave scattering length of \Rb{85} is near zero, so this cooling is again sympathetic \textendash\ although the isotope selectivity is less pronounced since both species see the same trapping potential.

After $5\units{s}$ of evaporation, the cloud is small enough to be contained entirely within the cross-beam of the dipole trap and the QUIC trap coils are switched off. At this point, the Feshbach field is slowly ramped from the inelastic minimum at $168\units{G}$ to near $163\units{G}$, where the \Rb{85} scattering length is $a=+200\units{$a_0$}$, suitable for the formation of a stable condensate. With a further $2\units{s}$ of evaporation at this field, we observe Bose-Einstein condensation of \Rb{85} (Figure \ref{fig:formation}). The final trapping frequencies of the crossed dipole trap are $\omega_{x,y,z} = 2\pi\times(50,57,28)\units{Hz}$. By varying the relative number of \Rb{85} and \Rb{87} initially loaded in the MOTs we can optimize for maximum final phase space density in \Rb{85}, and choose whether to have a \Rb{87} BEC of up to $10^6$ atoms coexisting in the trap.

We typically reach the phase transition with $3\times10^5$ \Rb{85} atoms, and can create nearly pure condensates containing $4\times10^4$ atoms with a peak density of $2\times10^{13}\units{cm$^{-3}$}$. The presence of a BEC is clearly indicated by the emergence of a bimodal density distribution and by the reversal of the aspect ratio of the cloud during expansion. The lifetime of the condensate in the crossed dipole trap is around $3\units{s}$.

When creating dual-species condensates, we also observe phase separation of the two isotopes due to the large interspecies scattering length $a_{85-87}=+213\units{$a_0$}$ \cite{burke99}. At fields where the interspecies repulsion is greater than the intraspecies repulsion, the two fluids will be immiscible and the condensates will spatially separate. Because the \Rb{85} scattering length may be changed using the Feshbach resonance, a dual-species \Rb{85}\textendash\Rb{87} BEC exhibits tunable miscibility, an effect first observed by Papp \etal\ \cite{papp08}. The miscibility parameter $\mu$ is defined as:
\begin{equation}
\mu = \frac{a_{85} a_{87}}{a^2_{85-87}} \quad.
\end{equation}
For the case of \Rb{85} and \Rb{87}, when $a_{85}>460\units{$a_0$}$, $\mu>1$ and the condensates are miscible. In our experiment, \Rb{85} condensates are created with a scattering length $a=+200\units{$a_0$}$, corresponding to the immiscible regime when both species coexist in the trap. Figure \ref{fig:immiscibility} shows the density distribution of a partly condensed \Rb{85} cloud in the presence of a large \Rb{87} condensate. The thermal \Rb{85} component is spread symmetrically over the trapping volume, while the condensed fraction is repelled by the \Rb{87} condensate and forms a crescent-shaped cloud at the edge of the trap.

\fig{0.23}{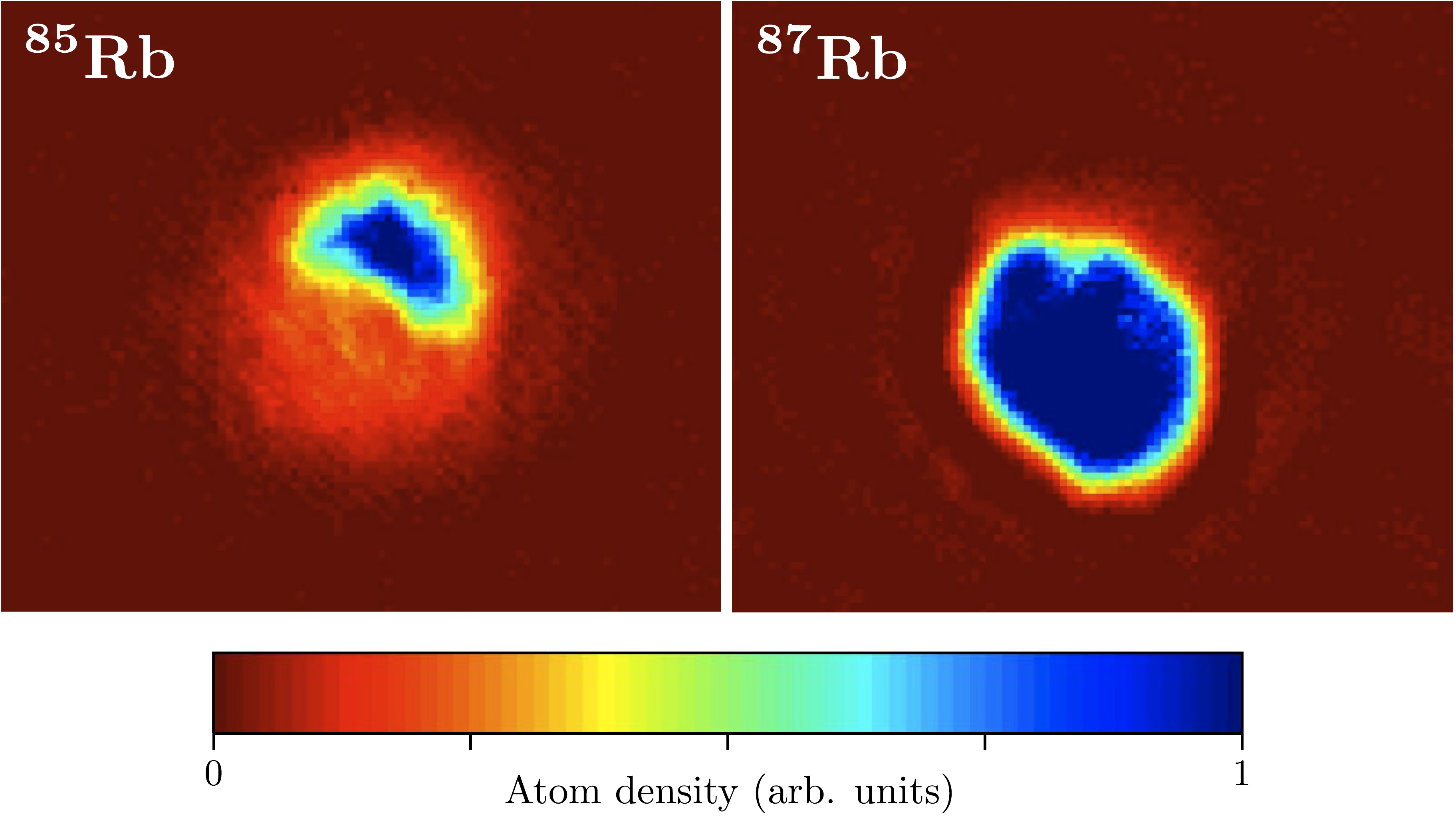}{Phase separation of \Rb{85} and \Rb{87} condensates in the immiscible regime. At $a_{85}=+200\units{$a_0$}$, the interspecies repulsion exceeds the intraspecies repulsion (miscibility parameter $\mu<1$) and the clouds separate, producing a small, crescent-shaped \Rb{85} condensate at the edge of a larger \Rb{87} cloud.}

\section{Comparison}

The first machine to make a \Rb{85} Bose-Einstein condensate with tunable interactions was built at JILA, Colorado in 2000 \cite{cornish00}. It utilized careful and lengthy direct evaporation of \Rb{85} in a weak magnetic trap with a large and variable bias field, and produced condensates containing up to $16,000$ atoms. In 2006, the same group completed a new machine for making \Rb{85} BECs by sympathetic cooling with \Rb{87}, producing larger condensates (up to $80,000$ atoms) with a shorter duty cycle \cite{papp07}. It is on this latest work that our method is largely based. However, although the procedure is similar, our setup has several important simplifications and a significantly shorter duty cycle.

Our vacuum system is more compact than that used in the work of Papp \etal\ \cite{papp07}, at the expense of a reduced background lifetime due to the science cell being in direct line-of-sight of the 2D MOT. However, this is compensated by speedier evaporation. The number of atoms collected in the 3D MOT is several times larger in our machine, but we do not perform any spin-polarization, resulting in a similar number of atoms being loaded into the magnetic trap. Although our apparatus requires a 2D MOT in addition to the 3D MOT, it enables a very simple single-coil magnetic transfer from the MOT to the Ioffe-Pritchard trap. This is compared to the set of ten coils and mechanical translation stage in the JILA machine, creating a large set of parameters to be optimized to ensure efficient transport with minimal heating. The sympathetic cooling in the magnetic trap follows a qualitatively similar trajectory in the two machines, including the efficiency roll-off as the number of each species becomes comparable. Two notable differences are the initial temperature \textendash\ $200\units{$\mu$K}$ here compared to $700\units{$\mu$K}$ at JILA, leading to a higher final phase space density in our machine \textendash\ and the length of the rf evaporation stage; 15\units{s} in this work compared to $44\units{s}$.

The setup for our optical trap is greatly simplified compared to that at JILA as it does not require active stabilisation, which also means that we can utilize the full power of our fibre laser. Also, our much larger beam waists ($150-200\units{$\mu$m}$ compared with $46\units{$\mu$m}$) remove the need for high quality, diffraction-limited optics. Despite lower trapping frequencies (our initial $\omega_\rho$ is an order of magnitude lower than in Ref \cite{papp07}), we evaporate to BEC in $7\units{s}$, compared to $11.5\units{s}$ in the JILA machine. An entire run of the experiment takes only $30\units{s}$. Although we can create similar-sized \Rb{87} condensates of around $10^6$ atoms, our \Rb{85} BECs are about half the size of the largest created in the JILA machine, and we find the optimum scattering length for making BEC to be $a=+200\units{$a_0$}$, rather than $a=60\units{$a_0$}$ as found by Papp \etal\ Importantly, our final samples are confined purely optically, without an axial field curvature as in Ref \cite{papp07}, making possible experiments with spinor \Rb{85} BECs.

\section{Conclusion and Outlook}

We have presented details of our machine for creating Bose-Einstein condensates of \Rb{85} with tunable interactions, which has only been achieved by one other group. We employ sympathetic cooling with \Rb{87}, first in a magnetic trap and then in a crossed optical dipole trap, to make \Rb{85} BECs containing $4\times10^4$ atoms. Our apparatus has several important simplifications compared with the \Rb{85} BEC machine at JILA. Although our \Rb{85} BECs are currently smaller than those created in that machine, we have thus far seen increases in the number of atoms in the Ioffe-Pritchard trap translate directly into larger condensates, and expect to be able to increase this number by spin-polarizing the gas prior to loading the magnetic trap. The machine presented here confines \Rb{85} BECs in a purely optical potential and is also able to produce large \Rb{87} condensates, giving us the flexibility to perform a variety of experiments involving dual-species and spinor BECs.

\section*{ACKNOWLEDGMENTS}

The authors gratefully acknowledge B. Buchler for experimental assistance, G. Dennis and J. Hope for help with numerical simulations, and N. Hinchey, S. Grieves, P. Tant, P. McNamara and N. Devlin for technical assistance. This work is supported by the Australian Research Council Centre of Excellence for Quantum-Atom Optics.

\end{document}